\documentclass[twocolumn,preprintnumbers,amsmath,amssymb]{revtex4}

\usepackage{graphicx}
\usepackage{amsmath}

\usepackage{xspace}

\begin{document}

\title{Improvement of the poly-3-hexylthiophene transistor performance using small molecule contact functionalization}

\author{Rebecca Winter $^1$}
\email[Electronic mail: ]{rebecca.winter@physik.uni-wuerzburg.de}
\author{Maria S Hammer $^1$}
\author{Carsten Deibel $^1$}
\author{Jens Pflaum $^{1,2}$}
\email[Electronic mail: ]{jpflaum@physik.uni-wuerzburg.de}
\affiliation{$^1$ Experimental Physics VI, Faculty of Physics and Astronomy, Julius-Maximilians-University of W\"urzburg, Am Hubland, 97074 W\"urzburg, Germany}
\affiliation{$^2$Bavarian Center of Applied Energy Research (ZAE Bayern), Am Hubland, 97074 W\"urzburg, Germany}

\begin{abstract}
We demonstrate an approach to improve poly-3-hexylthiophene field effect transistors by modifying the gold contacts with monolayer thick pentacenequinone (PQ) or naphthalene (NL). The effective contact resistance is reduced by a factor of two and sixteen for interlayers of PQ and NL, respectively. The observation is attributed to different injection barriers at the metal-organic interface caused by the functionalization and to an additional tunneling barrier enhancing the on/off ratios. This barrier yields to activation energies of 37meV (NL) and 104meV (PQ) below 190K, which are smaller than without functionalization, 117meV.
\end{abstract}

\maketitle

\clearpage
In recent years polymer electronics came into the focus of research due to their low cost potential and solution processability. In this context, poly-(3-hexylthiophene) (P3HT) has become one of the prototypical candidates for device applications due to its high field effect mobilities for holes of up to 10$^{-1}$cm$^2$V$^{-1}$s$^{-1}$ \cite{sirringhaus1999}. The charge transport mechanism is ascribed to hopping of charge carriers between localized states \cite{baessler1993, barankovskii2000} resulting in rather low mobilities, in contrast to inorganic semiconductors. Additionally, P3HT displays a high sensitivity on the $\pi$--$\pi$ stacking of the hexylthiophene chains \cite{zen2004}. 

Though many efforts have been spent on improving the effective hole mobility in P3HT films, less activities address the issue of charge injection at the metal--polymer interface~\cite{arkhipov1999a, wolf1999, scott1999a}. The underlying mechanisms turn out to be very complex and cannot simply be described by thermionic emission or tunneling through a potential barrier, but rather by thermally assisted tunneling from delocalized metal states into the spatially and energetically distributed localized states of the polymeric semiconductor \cite{scott2003}. To achieve high performance P3HT field effect transistors (FETs) the optimization of the contact resistance is inevitable.

We modified the metal--polymer interface by adsorbing monolayers of small polyaromatic molecules. Similar approaches using alkanethiole monolayers chemisorbed on Au contact structures have been reported in literature~\cite{marmont2008, cai2008}. However these self--assembled interlayers have the drawback of long chains, yielding rather thick tunneling barriers. In our approach, we used the oligoacenes pentacenequinone (PQ) and naphthalene (NL) each having the $\pi$-electron system delocalized over two aromativ rings, in combination with Au bottom contacts. As both molecular species show large ionization energies (IE) of 6.4~eV (PQ) and 8.15~eV(NL) \cite{watkins2002, koch2006, biermann1980}, they form tunneling barriers for hole injection into the P3HT highest-occupied molecular orbital (HOMO) localized at IE~=~4.9~eV \cite{chua2005}. Moreover as confirmed by tunneling microscopy studies, such polyaromatic entities prefer a flat-lying orientation on top of metal surfaces, i.e. the barrier of our functionalized contacts is about 3~\AA~thick. Finally, photo-electron spectroscopy data on PQ on Ag have shown that the HOMO position of a subsequent oligoacene layer, in this case pentacene can be substantially shifted by about 0.7~eV towards the metal Fermi-level \cite{koch2006}.

We observe that at room temperature transistors with contact modification by PQ and NL exhibit smaller contact resistances and accordingly larger injection currents than transistors without functionalization (WF). Temperature dependent measurements revealed a thermally activated contribution to the injection which can be further discriminated by a low and a high temperature regime. We discuss possible origins of our findings in the context of the formation of interface dipoles and thermally induced conduction state narrowing in the polymeric transport layer.

\begin{figure} 
  \centering
 \includegraphics[width=0.8\linewidth]{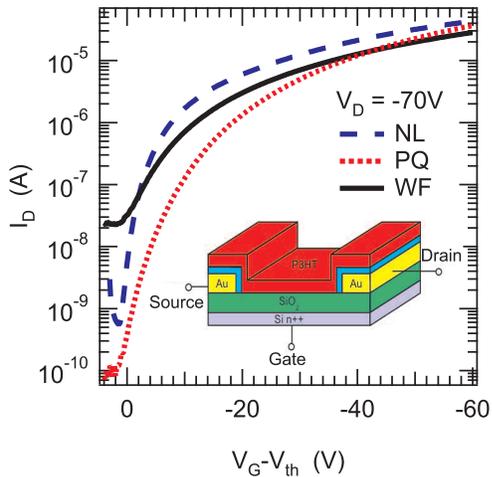} 
  \caption{(Color online) Transfer characteristic for fets with each $L~=~40\mu$m and $W~=~96mm$ for different functionalizations as indicated by the legend. Inset: The basic bottom contact structure with functionalization (blue interlayer).}
  \label{fig:transsat40}
\end{figure}

The FETs, sketched in Figure~\ref{fig:transsat40}, consist of a 200~nm thick SiO$_2$ dielectric thermally grown on top of highly n--doped Si(001). Interdigitated bottom contact structures were prepared by photolithography with channel lengths $L$ between 10~$\mu$m and 320~$\mu$m and widths $W$ between 37.5~mm and 119~mm. The thermally evaporated contacts consist of 19~nm Au on-top of a 1~nm thick Ti adhesion layer. We completed standard cleaning procedures by ultrasonification in acetone and isopropanol. Contact functionalization by PQ has been carried out by sublimation of 2~nm PQ under high vacuum (10$^{-7}$~mbar) at substrate temperatures of 80$^\circ$~C. During the subsequent annealing step at 140$^\circ$~C for 2~hours most of the PQ was desorbed from the SiO$_2$ and the Au surfaces as confirmed by thermal desorption spectroscopy (TDS). Furthermore, as a result from normalized TDS, we expect an interfacial layer of PQ to remain on the Au contacts due to the polarizability of the metal surface.

\begin{figure} 
  \centering
 \includegraphics[width=0.8\linewidth]{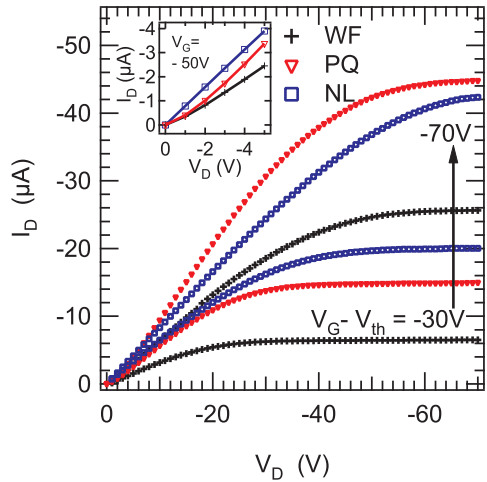} 
  \caption{(Color online) Output characteristics for fets with each $L~=~40\mu$m and $W~=~96mm$ for different functionalizations as indicated by the legend. Inset: output characteristics for  \mbox{V$_{G}$-V$_{th}$}=-50V at low $V_D$.}
  \label{fig:output}
\end{figure}

As the vapor pressure for NL at room temperature is high, simple exposure in a closed vessel for 18~hours is sufficient to deposit NL monolayers on the FET substrates. To desorb the NL multilayers the vessel with the samples was pumped to a pressure of 10$^{-3}$~mbar. Finally, all substrates were spin coated with 0.1~wt$\%$ rr-P3HT (98\% regio-regular; Rieke Metals, without further purification) solved in chlorobenzene at 3000~rpm for 60~s and subsequently annealed for 30~min at 80$^\circ$~C under nitrogen atmosphere.\\
As displayed in Fig.~\ref{fig:output} all FET devices show the expected output characteristics with an almost linear slope in the low voltage region and a saturation behavior for higher voltages. However, the source--drain current is  increased for the functionalized contact structures compared to that of the bare Au bottom contacts. Also, the functionalized FETs exhibit a more linear, i.e. ohmic, behavior at low $V_D$, whereas the FET without functionalization reveals a more pronounced s-shape, indicating an injection limited current (see inset in figure~\ref{fig:output}). $Vice~versa$ this implies that the effective contact resistance is reduced for the PQ and NL covered contacts. Comparing the output characteristics at different channel lengths (not shown), the relative improvement by functionalization is best for short $L$, since in short channels the contribution of the contact resistance to the total resistance of the FET is more pronounced. This observation is quite promising as short-channel transistors are the central building blocks for high frequency applications\cite{steudel2005}.

Assuming the channel resistance to be proportional to the length $L$, we extracted the contact resistance $R_C$ utilizing the transfer line method (TLM)~\cite{gundlach2006}. For this procedure, the total resistance at low $V_D$ is plotted versus $L$ for different $V_{G}$. The extrapolation to length zero reveals the bare contact resistance, which is almost independent of $V_{G}$. Table~\ref{tab:activation_energy} exemplarily shows the respective values at V$_{G}$=-50V for the different device structures at room temperature. Obviously, the chosen functionalization significantly reduces the contact resistance by a factor of 2 for PQ and by a factor of 16 for NL. From the slope of the length dependent total resistance we extracted the corrected mobility. The PQ and NL functionalization affects solely the film structure at the contact areas but not in the SiO$_2$ channel as can be concluded from the similar hole mobilities of about 10$^{-3}$~cm$^2$/Vs (see Table~\ref{tab:activation_energy}).

As an additional effect, the off-current for PQ and NL functionalization is found to be lower than for the bare P3HT-Au transistors (Fig.~\ref{fig:transsat40}). Therefore we obtain an increase of the on/off ratio by more than one order of magnitude for the NL and even two orders of magnitude for the PQ functionalized contacts compared to WF. We attribute the effect to the tunnel barrier imposed by the wide band gap oligoacene functionalization, which hampers charge carrier injection across the interface at small applied gate voltages.

\begin{figure} 
  \centering
 \includegraphics[width=0.8\linewidth]{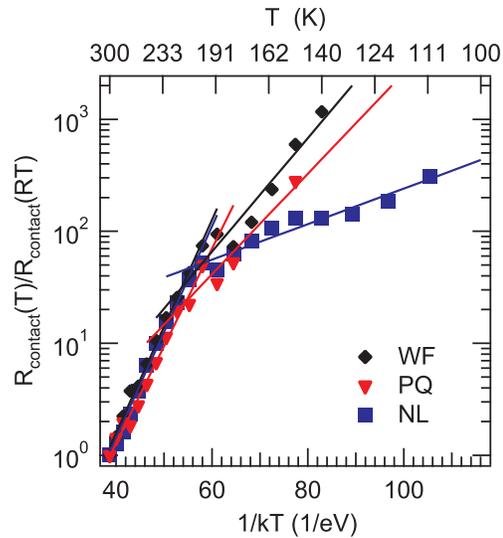} 
  \caption{(Color online) Temperature dependent data of the contact resistance at \mbox{$V_G-V_{th}=-20V$} for samples with PQ and NL and WF, respectively, as indicated by the legend.}
\label{fig:temp_resistance}
\end{figure}

To elucidate possible microscopic mechanisms leading to the observed contact resistance of our P3HT devices, we performed temperature dependent FET measurements between 100~K and 300~K. Figure~\ref{fig:temp_resistance} shows the thermal behavior of the contact resistance normalized to their room temperature values.
We determined the contact resistance by an alternative method suggested by Horowitz~et~al.~\cite{horowitz1999} using the transconductance and the drainconductance. The advantage of this method compared to TLM is that only one channel length is required and therefore irreproducible exposure to air upon switching between FETs of different lengths is avoided. Extracted values for $T$=150K are listed in Table~\ref{tab:activation_energy}. From the Arrhenius-type diagrams, two regions with different thermal activation energies can be distinguished for all devices. A regime of high thermal activation energy between 300~K and 190~K is followed by a region of lower thermal activation. This temperature behavior is believed to originate from the subtle balance between injection and transport and will be addressed in further comparative studies on high mobility organics such as pentacene. 
The respective values for the activation energies are listed in Table~\ref{tab:activation_energy}. The activation energy in the high temperature part is identical for all FETs and can be attributed to a thermally activated hopping mechanism. At low temperature the activation energy for functionalized FET is lower than for bare contacts. This can be attributed to an injection dominated by the tunneling through the additional barrier. In this temperature region thermally activation gets underpart.
 Furthermore, the temperature behavior of the mobility shows a similar behavior as the contact resistance. This indicates unambiguously that the contact resistance is not only determined by the bare injection at the metal-polymer interface but also by the subsequent drift of the injected charge carriers. Therefore, similar activation energies are attributed to both processes.

\begin{table}[htb]
	\caption{Contact resistance $R_C$, mobility $\mu$, activation energies $E_A$ for field effect transistors with PQ, NL and without functionalization (WF).}
	\begin{center}
	\begin{ruledtabular}
	\begin{tabular}{lccccc}
	sample &	\multicolumn{2}{c}{$R_C$	($\Omega m$)}& $\mu$ (cm$^2$V$^{-1}$s$^{-1}$)&	\multicolumn{2}{c} {$E_A$ (meV)}  \\
	 &RT&150K& RT	& $T_{high}$ & $T_{low}$  \\
	 \hline
	WF &	16$\times 10^4$	&79$\times 10^7$&	1$\times 10^{-3}$& 219& 117 \\
	PQ &	7$\times 10^4$	&34$\times 10^7$&	1$\times 10^{-3}$& 200 &104 \\
	NL &	1$\times 10^4$	&6$\times 10^7$&	1$\times 10^{-3}$&  217 & 37 \\
	\end{tabular}
	\end{ruledtabular}
	\end{center}
	\label{tab:activation_energy}
\end{table}

This correlation between the resistance and mobility lead to a different view of the commonly accepted resistor model to describe the injection into FETs. The assumption of three $independent$ resistors, for source, channel and drain is not valid in any case and might lead to misestimations. To further examine the mobility dependence of the contact resistance, we fit our data to a model for the thermal injection current $I_{inj}$ suggested by Scott and Malliaras~\cite{scott1999a, hamadani2004} .
\begin{equation}
\frac{1}{R_{C}(T)}\propto I_{inj}(T)= K(T,F)\mu(T) \exp(-\frac{\Phi_B}{kT})
\label{scott}
\end{equation}
$K(F,T)$ is a factor which depends on the temperature $T$ and the electric field $F$. A double~logarithmic plot of the contact resistance $R_C$ $vs.$ the mobility $\mu$ using Eq.~\ref{scott} shows a linear behavior with slope 1 for a Schottky barrier of $\Phi_B$~=~0~eV. The slope differs from 1 if $\Phi_B$~$\neq$~0~eV. From our normalized data for $R_C~vs.~\mu$ (not shown) we can extract slopes nearly value of 1 for all three devices which indicate low Schottky barriers in all cases. 
By simulation of the Eq. \ref{scott} we were able to calculate the Schottky barrier to 83~meV for NL and 72~meV for PQ functionalization and 72~meV WF.

As a main result, we observe lower activation energies in the low temperature region for transistors with functionalized contacts. This can be explained by the dominating tunneling probability through the additional barrier and therefore a better injection behavior for low temperatures for functionalized FET due to a negligible temperature dependent of the tunneling current. The lower activation energies at low temperature can be attributed to a thermally assisted tunneling.

In summary, we demonstrated a possible route of improving the injection characteristics in P3HT FETs by means of PQ or NL functionalized Au bottom contacts. The higher source drain currents can be ascribed to an additional interface dipole caused by the polyaromatic interlayer \cite{UPS}. The temperature dependent measurements revealed lower activation energies for devices with functionalized contacts especially below 190~K. At second, we explained this observation to an higher tunneling probability through the additional interlayer.

Acknowledgments\\
The DFG is acknowledged for financial support (project PF385).


\clearpage


\begin{thebibliography}{12}

\expandafter\ifx\csname natexlab\endcsname\relax\def\natexlab#1{#1}\fi
\expandafter\ifx\csname bibnamefont\endcsname\relax
 \def\bibnamefont#1{#1}\fi
\expandafter\ifx\csname bibfnamefont\endcsname\relax
 \def\bibfnamefont#1{#1}\fi
\expandafter\ifx\csname citenamefont\endcsname\relax
 \def\citenamefont#1{#1}\fi
\expandafter\ifx\csname url\endcsname\relax
 \def\url#1{\texttt{#1}}\fi
\expandafter\ifx\csname urlprefix\endcsname\relax\def\urlprefix{URL }\fi
\providecommand{\bibinfo}[2]{#2}
\providecommand{\eprint}[2][]{\url{#2}}

\bibitem[{\citenamefont{Sirringhaus et~al.}(1999)\citenamefont{Sirringhaus, Brown, Friend, Nielsen, Bechgaard, Langeveld--Voss, Spiering, Janssen, Meijer, Herwig, de~Leeuw}}]{sirringhaus1999}
\bibinfo{author}{\bibfnamefont{H.}~\bibnamefont{Sirringhaus}},
\bibinfo{author}{\bibfnamefont{P.~J.}~\bibnamefont{Brown}},
\bibinfo{author}{\bibfnamefont{R.~H.}~\bibnamefont{Friend}},
\bibinfo{author}{\bibfnamefont{M.~M.}~\bibnamefont{Nielsen}},
\bibinfo{author}{\bibfnamefont{K.}~\bibnamefont{Bechgaard}},
\bibinfo{author}{\bibfnamefont{B.~M.~W.}~\bibnamefont{Langeveld--Voss}},
\bibinfo{author}{\bibfnamefont{A.~J.~H.}~\bibnamefont{Spiering}},
\bibinfo{author}{\bibfnamefont{R.~A.~J.}~\bibnamefont{Janssen}},
\bibinfo{author}{\bibfnamefont{E.~W.}~\bibnamefont{Meijer}},
\bibinfo{author}{\bibfnamefont{P.}~\bibnamefont{Herwig}},
\bibnamefont{and} \bibinfo{author}{\bibfnamefont{D.~M.}~\bibnamefont{de~Leeuw}}, \bibinfo{journal}{Nature} \textbf{\bibinfo{volume}{401}},
 \bibinfo{pages}{685} (\bibinfo{year}{1999}).


\bibitem[{\citenamefont{Baessler}(1993)\citenamefont{Baessler}}]{baessler1993}
\bibinfo{author}{\bibfnamefont{H.}~\bibnamefont{Baessler}},
\bibinfo{journal}{Phys. Stat. Sol. B} \textbf{\bibinfo{volume}{175}},
 \bibinfo{pages}{15} (\bibinfo{year}{1993}).

\bibitem[{\citenamefont{Barankovskii et~al.}(2000)\citenamefont{Baranovskii, Cordes, Hensel, and Leising}}]{barankovskii2000}
\bibinfo{author}{\bibfnamefont{S.~D.}~\bibnamefont{Baranovskii}},
\bibinfo{author}{\bibfnamefont{H.}~\bibnamefont{Cordes}},
\bibinfo{author}{\bibfnamefont{F.}~\bibnamefont{Hensel}},
\bibnamefont{and} \bibinfo{author}{\bibfnamefont{G.}~\bibnamefont{Leising}}, \bibinfo{journal}{Phys. Rev. B} \textbf{\bibinfo{volume}{62}},
 \bibinfo{pages}{7934} (\bibinfo{year}{2000}).


\bibitem[{\citenamefont{Zen et~al.}(2004)\citenamefont{Zen, Pflaum, Hirschmann, Zhuang, Jaiser, Asawapriom, Rabe, Scherf, and Neher}}]{zen2004}
\bibinfo{author}{\bibfnamefont{A.}~\bibnamefont{Zen}},
\bibinfo{author}{\bibfnamefont{J.}~\bibnamefont{Pflaum}},
\bibinfo{author}{\bibfnamefont{S.}~\bibnamefont{Hirschmann}},
\bibinfo{author}{\bibfnamefont{W.}~\bibnamefont{Zhuang}},
\bibinfo{author}{\bibfnamefont{F.}~\bibnamefont{Jaiser}},
\bibinfo{author}{\bibfnamefont{U.}~\bibnamefont{Asawapriom}},
\bibinfo{author}{\bibfnamefont{J.~P.}~\bibnamefont{Rabe}},
\bibinfo{author}{\bibfnamefont{U.}~\bibnamefont{Scherf}},
\bibnamefont{and} \bibinfo{author}{\bibfnamefont{D.}~\bibnamefont{Neher}}, \bibinfo{journal}{Adv. Funct. Mater.} \textbf{\bibinfo{volume}{14}},
 \bibinfo{pages}{757} (\bibinfo{year}{2004}).


\bibitem[{\citenamefont{Arkhipov et~al.}(1999)\citenamefont{Arkhipov, Wolf, and Baessler}}]{arkhipov1999a}
\bibinfo{author}{\bibfnamefont{V.~I.}~\bibnamefont{Arkhipov}},
\bibinfo{author}{\bibfnamefont{U.}~\bibnamefont{Wolf}},
\bibnamefont{and} \bibinfo{author}{\bibfnamefont{H.}~\bibnamefont{Baessler}}, \bibinfo{journal}{Phys. Rev. B} \textbf{\bibinfo{volume}{59}},
 \bibinfo{pages}{7414} (\bibinfo{year}{1999}).
	
\bibitem[{\citenamefont{Wolf et~al.}(1999)\citenamefont{Wolf, Arkhipov, and Baessler}}]{wolf1999}
\bibinfo{author}{\bibfnamefont{U.}~\bibnamefont{Wolf}},
\bibinfo{author}{\bibfnamefont{V.~I.}~\bibnamefont{Arkhipov}},
\bibnamefont{and} \bibinfo{author}{\bibfnamefont{H.}~\bibnamefont{Baessler}}, \bibinfo{journal}{Phys. Rev. B} \textbf{\bibinfo{volume}{59}},
 \bibinfo{pages}{7507} (\bibinfo{year}{1999}).

\bibitem[{\citenamefont{Scott et~al.}(1999)\citenamefont{Scott, and Malliaras}}]{scott1999a}
\bibinfo{author}{\bibfnamefont{J.~C.}~\bibnamefont{Scott}}
\bibnamefont{and} \bibinfo{author}{\bibfnamefont{G.~G.}~\bibnamefont{Malliaras}}, \bibinfo{journal}{Chem. Phys. Lett.} \textbf{\bibinfo{volume}{299}},
 \bibinfo{pages}{115} (\bibinfo{year}{1999}).

\bibitem[{\citenamefont{Scott}(2003)}]{scott2003}
\bibinfo{author}{\bibfnamefont{J.~C.}~\bibnamefont{Scott}},
\bibinfo{journal}{J. Vac. Sci. Technol. A} \textbf{\bibinfo{volume}{21}},
 \bibinfo{pages}{521} (\bibinfo{year}{2003}).

\bibitem[{\citenamefont{Cai et~al.}(2008)\citenamefont{Cai, Chan-Park, Zhou, Lu, Li, and Ong}}]{cai2008}
\bibinfo{author}{\bibfnamefont{Q.~J.}~\bibnamefont{Cai}},
\bibinfo{author}{\bibfnamefont{M.~B.}~\bibnamefont{Chan-Park}},
\bibinfo{author}{\bibfnamefont{Q.}~\bibnamefont{Zhou}},
\bibinfo{author}{\bibfnamefont{Z.~S.}~\bibnamefont{Lu}},
\bibinfo{author}{\bibfnamefont{C.~M.}~\bibnamefont{Li}},
\bibnamefont{and} \bibinfo{author}{\bibfnamefont{B.~S.}~\bibnamefont{Ong}}, \bibinfo{journal}{Org. Electron.} \textbf{\bibinfo{volume}{9}},
 \bibinfo{pages}{936} (\bibinfo{year}{2008}).

\bibitem[{\citenamefont{Marmont et~al.}(2008)\citenamefont{Marmont, Battaglini,Lang, Horowitz, Hwang, Kahn, Amato, and Calas}}]{marmont2008}
\bibinfo{author}{\bibfnamefont{P.}~\bibnamefont{Marmont}},
\bibinfo{author}{\bibfnamefont{M.~B.}~\bibnamefont{Battaglini}},
\bibinfo{author}{\bibfnamefont{Q.}~\bibnamefont{Lang}},
\bibinfo{author}{\bibfnamefont{Z.~S.}~\bibnamefont{Horowitz}},
\bibinfo{author}{\bibfnamefont{C.~M.}~\bibnamefont{Hwang}},
\bibinfo{author}{\bibfnamefont{C.~M.}~\bibnamefont{Kahn}},
\bibinfo{author}{\bibfnamefont{C.~M.}~\bibnamefont{Amato}},
\bibnamefont{and} \bibinfo{author}{\bibfnamefont{B.~S.}~\bibnamefont{Calas}}, \bibinfo{journal}{Org. Electron.} \textbf{\bibinfo{volume}{9}},
 \bibinfo{pages}{419} (\bibinfo{year}{2008}).


\bibitem[{\citenamefont{Watkins et~al.}(2002)\citenamefont{Watkins, Yan, and Gao}}]{watkins2002}
\bibinfo{author}{\bibfnamefont{N.~J.}~\bibnamefont{Watkins}},
\bibinfo{author}{\bibfnamefont{L.}~\bibnamefont{Yan}},
\bibnamefont{and} \bibinfo{author}{\bibfnamefont{Y.~L.}~\bibnamefont{Gao}}, \bibinfo{journal}{Appl. Phys. Lett.} \textbf{\bibinfo{volume}{80}},
 \bibinfo{pages}{4384} (\bibinfo{year}{2002}).

\bibitem[{\citenamefont{Koch et~al.}(2006)\citenamefont{Koch, Salzmann, Johnson, Pflaum, Friedlein, and Rabe}}]{koch2006}
\bibinfo{author}{\bibfnamefont{N.}~\bibnamefont{Koch}},
\bibinfo{author}{\bibfnamefont{I.}~\bibnamefont{Salzmann}},
\bibinfo{author}{\bibfnamefont{R.~L.}~\bibnamefont{Johnson}},
\bibinfo{author}{\bibfnamefont{J.}~\bibnamefont{Pflaum}},
\bibinfo{author}{\bibfnamefont{R.}~\bibnamefont{Friedlein}},
\bibnamefont{and} \bibinfo{author}{\bibfnamefont{J.~P.}~\bibnamefont{Rabe}}, \bibinfo{journal}{Org. Electron.} \textbf{\bibinfo{volume}{7}},
 \bibinfo{pages}{537} (\bibinfo{year}{2006}).

\bibitem[{\citenamefont{Biermann et~al.}(1980)\citenamefont{Biermann, and Schmidt}}]{biermann1980}
\bibinfo{author}{\bibfnamefont{D.}~\bibnamefont{Biermann}},
\bibnamefont{and} \bibinfo{author}{\bibfnamefont{W.}~\bibnamefont{Schmidt}}, \bibinfo{journal}{J. Am. Chem. Soc.} \textbf{\bibinfo{volume}{102:9}},
 \bibinfo{pages}{3163} (\bibinfo{year}{1980}).

\bibitem[{\citenamefont{Chua et~al.}(2005)\citenamefont{Chua, Zaumseil, Chang, Ou, Ho, Sirringhaus and Friend}}]{chua2005}
\bibinfo{author}{\bibfnamefont{L.~L.}~\bibnamefont{Chua}},
\bibinfo{author}{\bibfnamefont{J.}~\bibnamefont{Zaumseil}},
\bibinfo{author}{\bibfnamefont{J.~F.}~\bibnamefont{Chang}},
\bibinfo{author}{\bibfnamefont{E.~C.-W.}~\bibnamefont{Ou}},
\bibinfo{author}{\bibfnamefont{P.~K.-H.}~\bibnamefont{Ho}},
\bibinfo{author}{\bibfnamefont{H.}~\bibnamefont{Stirringhaus}},
 \bibnamefont{and} \bibinfo{author}{\bibfnamefont{R.~H.}~\bibnamefont{Fiend}}, \bibinfo{journal}{Nature} \textbf{\bibinfo{volume}{434}},
  \bibinfo{pages}{194} (\bibinfo{year}{2005}).



\bibitem[{\citenamefont{Gundlach et~al.}(2006)\citenamefont{Gundlach, Zhou, Nochols, and Jackson, Necliudov, and Shur}}]{gundlach2006}
\bibinfo{author}{\bibfnamefont{D.~J.}~\bibnamefont{Gundlach}},
\bibinfo{author}{\bibfnamefont{L.}~\bibnamefont{Zhou}},
\bibinfo{author}{\bibfnamefont{J.~A.}~\bibnamefont{Nichols}},
\bibinfo{author}{\bibfnamefont{T.~N.}~\bibnamefont{Jackson}},
\bibinfo{author}{\bibfnamefont{P.~V.}~\bibnamefont{Necliudov}},
\bibnamefont{and} \bibinfo{author}{\bibfnamefont{M.~S.}~\bibnamefont{Shur}}, \bibinfo{journal}{J. Appl. Phys.} \textbf{\bibinfo{volume}{100}},
 \bibinfo{pages}{024509} (\bibinfo{year}{2006}).

\bibitem[{\citenamefont{Hamadani et~al.}(2004)\citenamefont{Hamadani, and }}]{hamadani2004}
\bibinfo{author}{\bibfnamefont{B.~H.}~\bibnamefont{Hamadani}},
\bibnamefont{and} \bibinfo{author}{\bibfnamefont{D.}~\bibnamefont{Natelson}}, \bibinfo{journal}{Appl. Phys. Lett.} \textbf{\bibinfo{volume}{84}},
 \bibinfo{pages}{443} (\bibinfo{year}{2004}).

\bibitem[{\citenamefont{Horowitz et~al.}(1999)\citenamefont{Horowitz, Hajlaoui, Fichou,  and Kassmi}}]{horowitz1999}
\bibinfo{author}{\bibfnamefont{G.}~\bibnamefont{Horowitz}},
\bibinfo{author}{\bibfnamefont{R.}~\bibnamefont{Hajlaoui}},
\bibinfo{author}{\bibfnamefont{D.}~\bibnamefont{Fichou}},
\bibnamefont{and} \bibinfo{author}{\bibfnamefont{A.}~\bibnamefont{El~Kassmi}}, \bibinfo{journal}{J. Appl. Phys.} \textbf{\bibinfo{volume}{85}},
 \bibinfo{pages}{3202} (\bibinfo{year}{1999}).


\bibitem[{\citenamefont{Steudel et~al.}(2005)\citenamefont{Steudel, Myny, Arkhipov, Deibel,  de Vusser, Genoe, and Heremans}}]{steudel2005}
\bibinfo{author}{\bibfnamefont{S.}~\bibnamefont{Steudel}},
\bibinfo{author}{\bibfnamefont{R.}~\bibnamefont{Myny}},
\bibinfo{author}{\bibfnamefont{D.}~\bibnamefont{Arkhipov}},
\bibinfo{author}{\bibfnamefont{C.}~\bibnamefont{Deibel}},
\bibinfo{author}{\bibfnamefont{S.}~\bibnamefont{de~Vusser}},
\bibinfo{author}{\bibfnamefont{J.}~\bibnamefont{Genoe}},
\bibnamefont{and} \bibinfo{author}{\bibfnamefont{P.}~\bibnamefont{Heremans}}, \bibinfo{journal}{Nat. Mater.} \textbf{\bibinfo{volume}{4}},
 \bibinfo{pages}{597} (\bibinfo{year}{2005}).


\bibitem{UPS}
\bibinfo{note}{Preliminary UPS measurements on the material systems PQ/Au/P3HT and Au/P3HT indicate a larger interface dipole in the case of the former.}.
\end{thebibliography}
\end{document}